\documentclass[%
superscriptaddress,
 prl,
rsi,%
amsmath,amssymb,
reprint,%
floatfix,%
]{revtex4-1}

\usepackage{graphicx}
\usepackage{dcolumn}
\usepackage{xcolor}
\usepackage{float}

\usepackage{xspace}
\usepackage{amsmath}

\usepackage{multirow}
\usepackage{array}

\usepackage{lipsum}
\usepackage{subfigure}
\usepackage[justification=raggedright]{caption}

\begin{document}
\author{Cl\'ement Guiot du Doignon}
\affiliation{ 
Université Claude Bernard Lyon 1, CNRS, Institut Lumière Matière, UMR5306, F-69100 Villeurbanne, France
}%
\author{Rajarshi Sinha-Roy}%
\affiliation{ 
Université Claude Bernard Lyon 1, CNRS, Institut Lumière Matière, UMR5306, F-69100 Villeurbanne, France
}%
\author{Franck Rabilloud}%
\affiliation{ 
Université Claude Bernard Lyon 1, CNRS, Institut Lumière Matière, UMR5306, F-69100 Villeurbanne, France
}%
\author{Victor Despr\'e}%
\email{victor.despre@univ-lyon1.fr}
\affiliation{ 
Université Claude Bernard Lyon 1, CNRS, Institut Lumière Matière, UMR5306, F-69100 Villeurbanne, France
}%

\title{Correlation-Driven Charge Migration Triggered by Infrared Multi-Photon Ionization}

\date{\today}

\begin{abstract}
 The possibility of observing correlation-driven charge migration has been a driving force behind theoretical and experimental developments in the field of attosecond molecular science since its inception. Despite significant accomplishments, the unambiguous experimental observation of this quantum beating remains elusive. In this work, we present a method to selectively trigger such dynamics using molecules predicted to exhibit long-lived electron coherence. We show that these dynamics can be selectively triggered using infrared multi-photon ionization and probed using the spacial resolution of X-ray free-electron laser, proposing a promising experimental scheme to study these pivotal dynamics. Additionally, we demonstrate that real-time time-dependent density-functional theory can describe correlation-driven charge migration resulting from a hole mixing structure involving the HOMO of a molecule.
\end{abstract}

\maketitle

The development of ultrafast technologies, which permit the study of matter down to the attosecond (as) time scale ($10^{-18}$) \cite{antoine1996attosecond,krausz2009attosecond}, has enabled the investigation of fundamental quantum effects that were previously inaccessible \cite{lepine2014attosecond,nisoli2017attosecond}. In molecular physics, this was first demonstrated through experiments on the fragmentation of $H_2$ \cite{sansone2010electron} and the ionization of small polyatomic molecules \cite{neidel2013probing}. Currently, the attosecond community aims to expand the application of ultrafast technologies into new research fields. For example, nowadays it has become possible to use extreme ultraviolet (XUV) pulses as a trigger on biologically relevant systems~\cite{calegari2014ultrafast}. Also its use on carbon based structures like polycyclic aromatic hydrocarbons is shedding new light on questions relevant to astrochemistry \cite{marciniak2019electron}. 

These advancements have often been driven and supported by new theoretical developments and ideas. One significant theoretical prediction that propelled the early development of attosecond technologies is the correlation-driven charge migration \cite{kuleff2014ultrafast}, as predicted by Lorenz Cederbaum \cite{cederbaum1999ultrafast}. Referred to as "the holy grail of attosecond molecular physics" \cite{lepine2013molecular}, charge migration involves ultrafast purely electron dynamics resulting from the coherent superposition of eigenstates in a molecular system. Correlation-driven charge migration specifically refers to systems where such a superposition can be created by ionizing a single molecular orbital, enabled by electron correlation \cite{breidbach2003migration}. Studying correlation-driven charge migration offers a direct and rich way to investigate electron correlation, one of the primary aims of attosecond technologies. 

However, despite significant theoretical \cite{kuleff2010ultrafast,kuleff2016core,mignolet2014charge,sun2017nuclear,lara-astiaso2016decoherence,yuan2019ultrafast,folorunso2023attochemistry,haase2021electron,zhang2024cavity} and experimental \cite{calegari2014ultrafast,lara-astiaso2018attosecond,kraus2015measurement,schwickert2022electronic} efforts, the unambiguous observation of correlation-driven charge migration, or more generally, charge migration dynamics triggered by ionization, remains elusive. Recently, X-ray free-electron lasers have emerged as a particularly promising tool for experimentally observing correlation-driven charge migration \cite{li2024attosecond,guo2024experimental,driver2024attosecond}. These large-scale infrastructures offer the exciting possibility of probing the charge distribution within a molecule at the atomic level by exploiting the spacial resolution of X-ray photons, potentially making the experimental observation of this long-sought-after phenomenon a reality.

Pursuing such an observation is crucial as charge migration is central to extending attosecond technologies into chemistry, a field referred to as attochemistry. The goal is to steer the chemical reactivity of a molecular system by controlling its purely electron dynamics, the charge migration. The study of attochemistry is timely, as two of its prerequisites have been demonstrated: the existence of long-lived electron coherence, as observed for neutral silane \cite{matselyukh2022decoherence}, and the impact of pure electron dynamics on the reactivity of a molecular system, as demonstrated for adenine \cite{maansson2021real,despre2022correlation}. In this context, ionization-triggered dynamics are particularly interesting as they allow control over the created hole in the system, directly influencing molecular reactivity \cite{remacle1998charge}.

A potential obstacle to observing ionization-induced charge migration is the decoherence of the superposition of states created during the process, first discussed in the context of the benzene molecule \cite{despre2015attosecond}. Predictions have been made for both long-lived coherence \cite{despre2015attosecond,despre2018charge} (around 15 fs, representing several periods of charge migration) and extremely short coherence (1-2 fs, which is only a fraction of the charge migration period) \cite{vacher2017electron,arnold2017electronic,vester2023role}. Long-lived coherence has predominantly been predicted for correlation-driven charge migration \cite{despre2015attosecond,despre2018charge,scheidegger2022search}. These studies have shown that selecting appropriate systems is crucial, but long-lived coherence is achievable. Current efforts aim to better understand which systems are likely to exhibit long-lived coherence \cite{vester2023role}, potentially leading to the design of molecules with interesting long-lived charge migration dynamics \cite{despre2019size,folorunso2021molecular,chordiya2023photo,belles2023size}. Despite identifying promising molecular systems, no direct experimental observation of correlation-driven charge migration has been made yet, raising questions about other possible barriers.

Experimental attempts to observe charge migration have mostly employed XUV-pump infrared (IR)-probe schemes \cite{calegari2014ultrafast,lara-astiaso2018attosecond}. However, the lack of selectivity of the pump could be a limiting factor. An XUV-pump will populate all cationic eigenstates of a molecule, as the pump spectrum usually extends beyond the double ionization threshold of most molecules. Consequently, the charge migration dynamics may not be efficiently populated or discernible from the experimental data. Nevertheless, the XUV-pump IR-probe scheme has achieved significant success, particularly in studying states just below the double ionization threshold of molecules \cite{marciniak2019electron,belshaw2012observation}, where the breakdown of the molecular picture \cite{cederbaum1986correlation} creates a correlation band \cite{deleuze1996formation,herve2021ultrafast,boyer2021ultrafast}.

In this work, we propose a different approach to triggering correlation-driven charge migration with strong selectivity. Multi-photon IR ionization is known to ionize a limited number of the outermost molecular orbitals, as demonstrated by above-threshold ionization experiments \cite{eberly1991above,milovsevic2006above}. Using intense IR pulses as a trigger will dramatically limit the number of populated cationic eigenstates. The question is how to trigger correlation-driven charge migration this way. We propose selecting molecular systems with a hole mixing structure for their HOMO. A hole mixing structure consists of two or more cationic states described by contributions from two or more different orbitals \cite{breidbach2003migration}. In the case of two states and two orbitals, ionizing one of the orbitals involved in the mixing will populate both states, with their populations determined by the corresponding weight of the orbitals in each state. Such hole mixing for the HOMO is not rare, as demonstrated by seminal work of the Heidelberg group on PENNA \cite{lunnemann2008ultrafast} and MePeNNA \cite{lunnemann2008charge}. Long-lived electron coherences have also been predicted for such structures, first for propiolic acid with a quantum treatment of both electronic and nuclear degrees of freedom \cite{despre2018charge}, coherence time that may even be controlled \cite{dey2022quantum}, and then for but-3-ynal, 2,5-dihydrofuran, and 3-pyrroline with a semi-classical approach \cite{scheidegger2022search}. These four molecules will be studied in this paper.

To simulate correlation-driven charge migration triggered by IR multiphoton ionization, the theoretical approach must meet several constraints. It must handle ionization and provide a high level of electron correlation treatment to describe hole mixing structures accurately. Real-time time-dependent density functional theory (RT-TDDFT) in its real-space formulation shows promise due to its unique ability to explicitly treat the ionization step, a capability lacking in the majority of other methods that rely on sudden ionization. RT-TDDFT was first shown to be relevant for attosecond science by simulating the control of XUV ionization induced by an IR polarizing pulse for $N_2$, $CO_2$, and $C_2H_4$ \cite{neidel2013probing}, and has since been used regularly in various studies  \cite{lucchini2016attosecond,tancogne2017impact,gao2017towards,bruner2017attosecond,hamer2024strong}. It handles ionization by adding absorbing boundaries at the simulation box's limits. While the ionization criterion is met, the question remains whether RT-TDDFT can properly describe correlation-driven charge migration. This has been addressed in the case of dynamics due to satellite states, where the single determinant nature of TDDFT led to nonphysical behavior due to the importance of multi-excitation for satellites \cite{kuleff2009theoretical}, which is not the case for hole mixing that are characterized by a sum of single excitation. Another consideration is how well Kohn-Sham orbitals describe ionization. Koopmans' theorem for the HOMO is favorable \cite{marques2006time}, but no such theorem exists for other orbitals.

In this paper, we first show that RT-TDDFT can accurately describe correlation-driven charge migration for propiolic acid, but-3-ynal, 2,5-dihydrofuran, and 3-pyrroline by simulating the sudden ionization of their HOMO. We then demonstrate that short intense IR pulses can selectively trigger these dynamics, proposing a novel approach for the experimental observation of correlation-driven charge migration.

\begin{figure*}
\begin{center}
\includegraphics[width=0.99\textwidth]{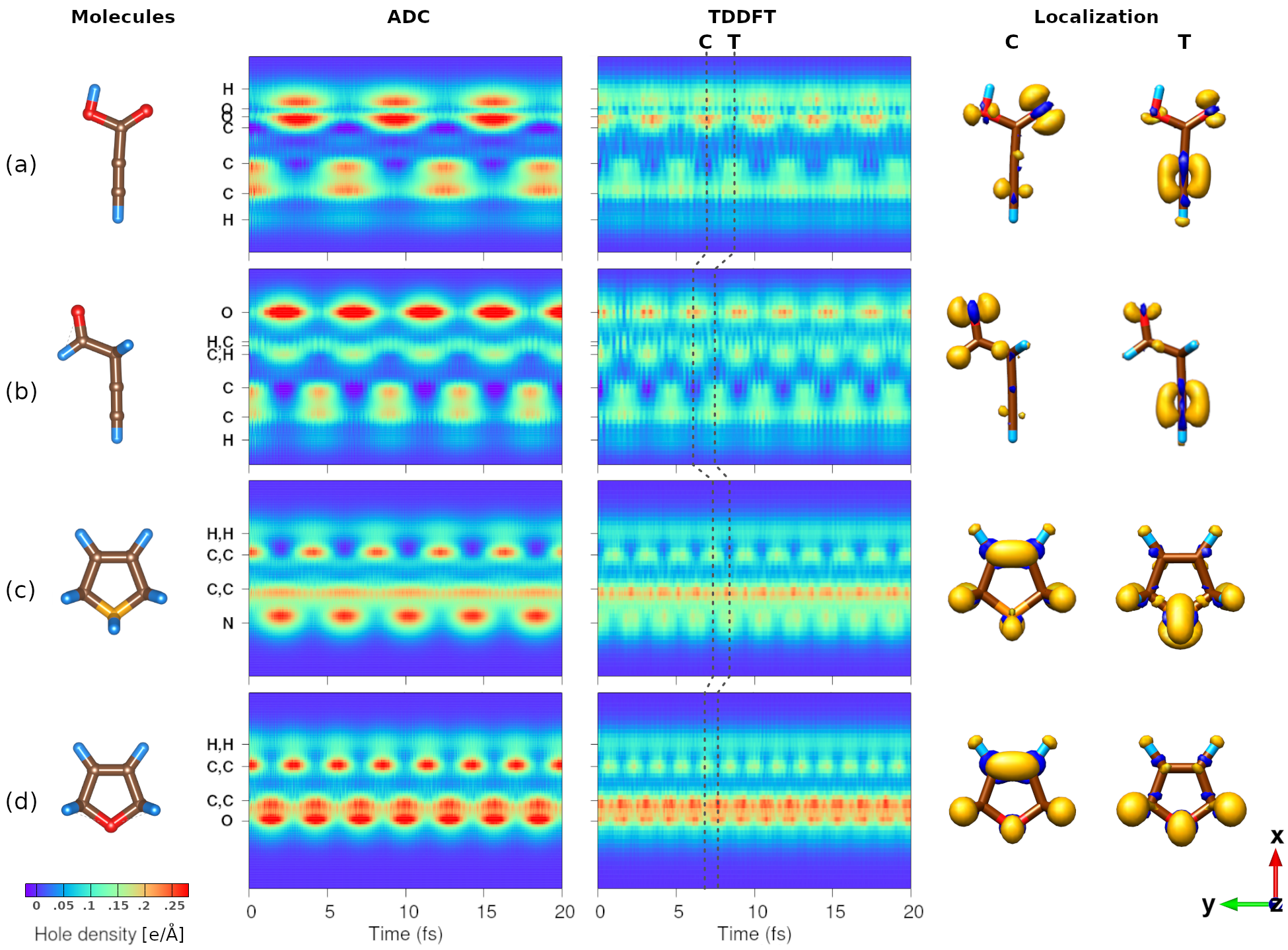}
\end{center}
\caption{The time-dependent oscillation of the hole density projected along the molecular axis due to the sudden ionization of the HOMO of propiolic acid (a), but-3-ynal (b), 3-pyrroline (c), and 2,5-dihydrofuran (d) is shown through color maps. 
The left-most column shows the molecules with their molecular axes oriented along the x-axis (cf., axes arrows).  
The third column represents calculations using TDDFT and compares to the results obtained in ADC as shown in the second one.
The values of the hole density are represented by the color bar. 
(A negative hole density means an excess of electron.) 
For the TDDFT results, the localization of the hole-density corresponding to a crests (C) and a trough (T) of the oscillatory dynamics is presented as iso-surface in the last two columns. The iso-values are $0.005 e/\AA^3$ for (a) and (b), and $0.008 e/\AA^3$ for (c) and (d).}
\label{sudden_ionization}
\end{figure*}

Simulations were performed using the open-source code \texttt{octopus}, which allows RT-TDDFT simulations in real space, i.e., on a grid \cite{andrade2015real,tancogne2020octopus}. This approach offers the advantage of straightforward convergence of the ionization process by increasing both the size of the simulation box and of the absorbing boundaries. All simulations presented in this paper are done with the PBE functional \cite{perdew1996generalized}. In addition, to correct the self-interaction error (SIE) so that the asymptotic behavior of the effective potential can be properly described, a scheme~\cite{legrand2002comparison,klupfel2013koopmans} based on the average density is used. The dependence of the predicted dynamics on the functional used will be discussed later. The  correction of the SIE significantly improves the ionization potential of the molecules, thereby enhancing the description of ionization. Simulations were conducted on a spherical grid with a radius of 12 Å, a spacing of 0.18 Å, and a time step of 1.3 as. Absorbing boundaries with a thickness of 2 Å were used at the edge of the sphere for simulation performed with the explicit inclusion of a laser pulse.

The first step of our study involved simulating correlation-driven charge migration resulting from the sudden ionization of the HOMO of the four molecules to evaluate the capability of RT-TDDFT simulations. To achieve this, an electron was removed from the HOMO of each molecule at the start of the simulation ($t=0$). The results, presented in Fig. 1, can be compared to dynamics predicted using the high-level {\it ab initio} method, the Algebraic Diagrammatic Construction at third order (ADC(3)) \cite{schirmer1982beyond,schirmer2018many}, as presented in ref \cite{despre2018charge} for propiolic acid and ref \cite{scheidegger2022search} for but-3-ynal, 2,5-dihydrofuran, and 3-pyrroline. In Fig. 1, the time-dependent hole density is projected along a molecular axis, as shown by the molecules presented on the left-most column, and the axes at the bottom-right corner of the figure. The hole density represents the difference between the electron density of the neutral system and the time-dependent electron density of the cationic system. Since the simulations do not account for nuclear dynamics or ionized electrons, decoherence is absent, and pure electron dynamics are maintained throughout.

Strong charge migration dynamics similar to the one predicted with ADC(3) \cite{despre2018charge,scheidegger2022search} is observed in all systems due to a hole mixing structure that allows the coherent population of several states through the ionization of a single molecular orbital. A similarity in the shape of the time-dependent hole density for the four molecules can be observed. For all systems, a major part of the dynamics involves the migration of charge between either an oxygen or nitrogen atom and a carbon-carbon triple or double bond. Periods of approximately 3.3 fs for propiolic acid, 2.9 fs for but-3-ynal, and 2 fs for 3-pyrroline and 1.55 fs for 2,5-dihydrofuran are predicted. These can be compared with periods of 6.2 fs for propiolic acid, 4.5 fs for but-3-ynal, and 4 fs for 3-pyrroline and 2.8 fs for 2,5-dihydrofuran predicted using ADC(3). The difference in periods calculated in the two different levels of theory is not surprising, it depends solely on the energy gap between the states in the coherent superposition, and these energies can vary depending on the level of theory used. In fact, a difference in the energy gap of just around 0.59 eV accounts for the largest observed difference in period, i.e., for the case of propiolic acid. Despite the difference in the period predicted by RT-TDDFT from that predicted by ADC(3), it is noteworthy how accurately
RT-TDDFT describes correlation-driven charge migration resulting from a hole mixing structure. It is crucial that RT-TDDFT accurately describes the distribution of the electron density, as this is the key element for studying charge dynamics. We want to mention that even the LDA functional correctly predicts the dynamics' shape. This demonstrates RT-TDDFT's ability to treat electron correlation sufficiently to describe hole mixing structures. 

The second question was how well Kohn-Sham orbitals describe ionization. The HOMO appears to be satisfactory, as expected from Koopmans' theorem. To test other orbitals, we simulated the sudden ionization of the second orbital involved in the hole mixing for each molecule. If Kohn-Sham orbitals accurately described ionization, the resulting dynamics would be similar to those obtained after ionizing the HOMO, which was not the case \cite{sinharoy2024tddft}. In the case of hole mixing between two orbitals, ionization from either orbital will result in very similar dynamics—exhibiting the same oscillation period but with opposite beating phases, meaning the initial and final charge localizations are reversed. Not observing this means that more care is needed when studying correlation effects during ionization for orbitals other than the HOMO. It is also worth noting that the initial charge localization differs between the ADC and RT-TDDFT simulations \cite{sinharoy2024tddft}.

\begin{figure}
\begin{center}
\includegraphics[width=8.5cm]{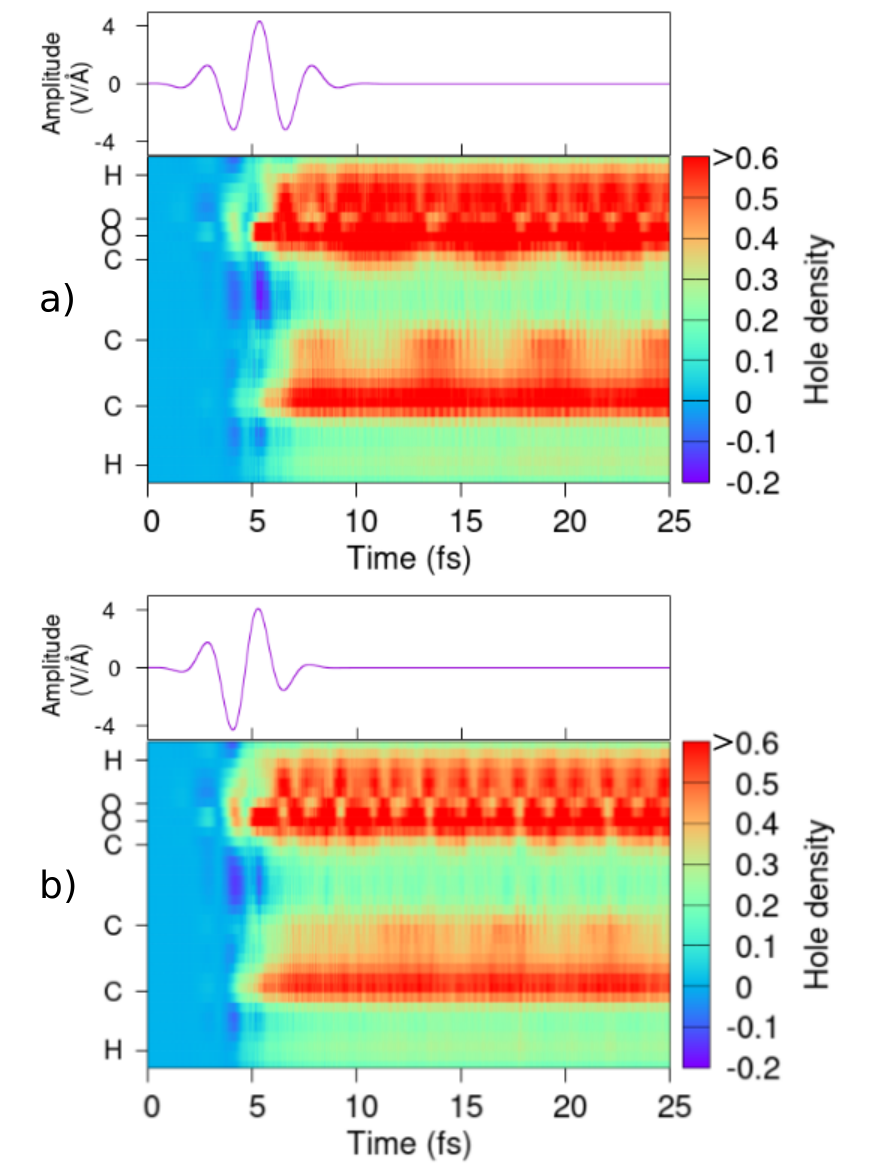}
\end{center}
\caption{Time-dependent hole density for propiolic acid projected along the molecular axis, shown as the y-axis of each panel, following multi-photon ionization triggered by an a) even or b) odd laser pulse, as illustrated above each panel. The hole density is normalized to its maximum value.}
\label{laser_ionization}
\end{figure}

The next step was to study how the predicted dynamics could be triggered by ionization using an intense IR pulse at 800 nm. To achieve this, a pulse was explicitly introduced into the simulations, which perturbs the ground states of the neutral molecules. A short pulse with a controlled carrier envelop phase (CEP) limits the time for which the intensity is sufficient for multi-photon ionization. The two pulses used are presented in Fig. 2, representing cases with one extremum for the field (even pulse) or two with equal intensity (odd pulse), both of which can be generated experimentally \cite{okell2013carrier}. In the following we will focus on propiolic acid, the only molecule in this study for which long-lived electron coherence has been predicted using a quantum treatment of both electronic and nuclear degrees of freedom \cite{despre2018charge} for a dynamics that can even be controlled by laser pulses \cite{golubev2017quantum}. The polarization of the field is chosen along the y axis as shown in Fig. 1 meaning that the polarization is perpendicular to the molecular axis. The intensity, around $2.5 \times 10^{14} W/cm^2$, was set to ensure a maximum ionization of 1. The level of ionization is determined by the difference between the initial charge within the simulation box and the final charge after part of the electron density passes through the absorbing boundaries. Dynamics are also observed with less intense pulses. As shown in Fig. 2, the same dynamics as the one obtained with sudden ionization is triggered for both pulses, indicating that the correlation-driven charge migration dynamics is efficiently and selectively triggered. However, the dynamics triggered by the odd pulse appear more blurred due to the significant population of higher-lying cationic states caused by the excitation of the created cation, meaning that coherent control occurs during the pulse. This secondary dynamics, which strongly localizes the hole around the singly bound oxygen, becomes dominant for certain orientations. Therefore, the even pulse appears more suitable for triggering the desired dynamics, indicating that the choice of the CEP is a crucial parameter for the experimental observation of these dynamics. Ultrafast beatings are also observed, resulting from the ionization of other orbitals, particularly HOMO-1 and HOMO-2, where sudden ionization leads to such patterns. This effect is partly due to the imperfect description of ionization by Kohn-Sham orbitals, resulting in a nonphysical superposition of states separated by a large energy gap. Longer pulses were tested and also triggered the desired dynamics; however, given that an electron coherence of approximately 15 fs is predicted for the studied molecule, longer pulses are less likely to enable experimental observation of the dynamics.

To guide experiments, we investigated whether molecular orientation is necessary for the experimental observation of the dynamics. We used the even pulse shown in Fig. 3, with an intensity around $1.6 \times 10^{14} W/cm^2$ and averaged the dynamics over 17 different molecular orientations. The choice of the intensity impact the contrast of the observed dynamics and must be chosen carefully. The chosen orientations included the three principal axes shown in Fig. 1 and their equal combinations, leveraging the molecule's $C_s$ symmetry, which allows us to explore only half of the physical space. The averaged dynamics are shown in Fig. 3a) for the full averaging and Fig. 3b) when only the 8 orientations lying in the plane of the molecule are considered. We found that the dynamics are still observed even with different orientations, though the contrast of spatial charge localization is better when only in-plane orientations are considered. This suggests that using an intense single-cycle IR pulse with a suitable CEP on randomly oriented molecules will selectively trigger the desired dynamics. While molecular alignment is not mandatory, it can enhance the observation of the dynamics by improving their charge-localization contrast.

\begin{figure}
\begin{center}
\includegraphics[width=8.5cm]{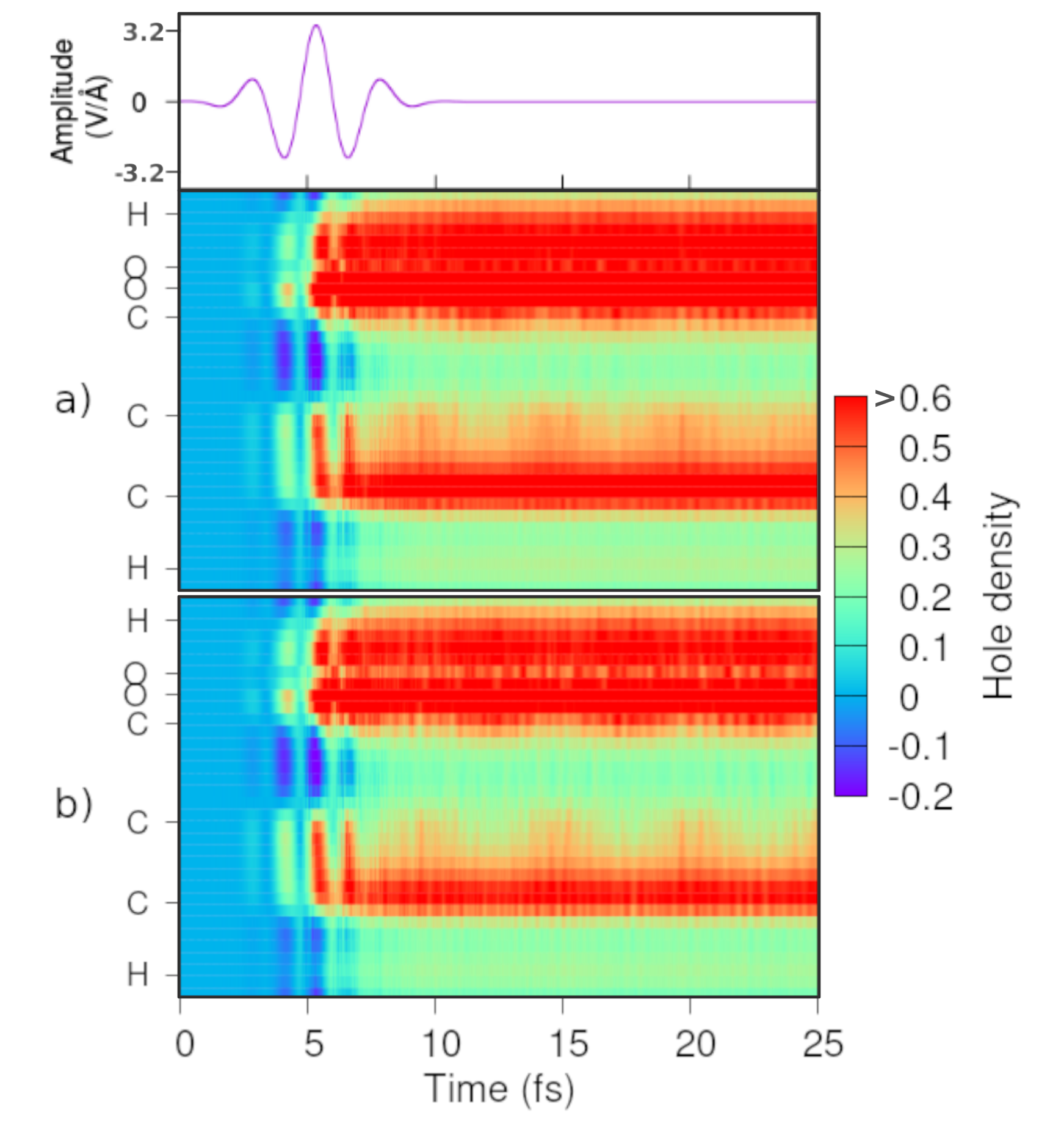}
\end{center}
\caption{Time-dependent hole density for propiolic acid projected along the molecular axis, shown as the y-axis of each panel, following multi-photon ionization triggered by the even pulse shown above the panels averaged over a) 17 and b) 8 orientation (see text). The hole density is normalized to its maximum value.}
\label{average_ionization}
\end{figure}

A key remaining question is how such dynamics can be effectively probed. One particularly attractive option is the use of X-ray free-electron lasers \cite{li2024attosecond,guo2024experimental,driver2024attosecond}, which have developed technologies capable of being highly sensitive to local processes by leveraging the spacial resolution of core transitions. Correlation-driven charge migration alters the screening of specific atoms within a molecule, leading to changes in core transitions. This enables the creation of a "molecular movie" \cite{yong2022attosecond}, allowing to track the movement of the hole within a molecule by probing different atoms. This approach has even been demonstrated theoretically for propiolic acid \cite{golubev2021core}. Consequently, an IR-pump/X-ray probe setup at an X-ray free-electron lasers facility appears to be a highly suitable scheme for studying correlation-driven charge migration.

In this paper, we have shown that RT-TDDFT can accurately describe correlation-driven charge migration dynamics resulting from hole mixing structures, even using the simplest functional. The Kohn-Sham HOMO adequately describes ionization, enabling the study of dynamics triggered by ionization with a short, intense IR pulse. We have demonstrated that these dynamics can be selectively triggered, with the choice of CEP and intensity playing a crucial role, while random molecular orientations can still be utilized. The resulting dynamics can then be effectively probed using X-ray free-electron lasers. We have proposed an efficient and selective alternative to using XUV attosecond pulses for triggering charge migration. While this approach is limited to systems with a hole mixing structure for the HOMO, such occurrences are not rare, and provide a direct and simple way to study electron coherence by unambiguously investigating the elusive correlation-driven charge migration mechanism. We hope our work will motivate new experiments aimed at observing charge migration dynamics and theoretical studies to deepen our understanding of the opportunities offered by RT-TDDFT in this context.

\begin{acknowledgments}
The authors thank Franck Lépine and Alexander Kuleff for fruitful discussions.
\end{acknowledgments}

\bibliography{RT-TDDFT.bib}

\end{document}